# Systems Biophysics of Gene Expression

Jose M. G. Vilar[1,2,*] and Leonor Saiz[3,*]
[1] Biophysics Unit (CSIC-UPV/EHU) and Department of Biochemistry and Molecular Biology, University of the Basque Country UPV/EHU, P.O. Box 644, 48080 Bilbao, Spain
[2] IKERBASQUE, Basque Foundation for Science, 48011 Bilbao, Spain
[3] Department of Biomedical Engineering, University of California, 451 E. Health Sciences Drive, Davis, CA 95616, USA

## Abstract

**Gene expression is a central process to any form of life. It involves multiple temporal and functional scales that extend from specific protein-DNA interactions to the coordinated regulation of multiple genes in response to intracellular and extracellular changes. This diversity in scales poses fundamental challenges among traditional approaches to fully understand even the simplest gene expression systems. Recent advances in computational systems biophysics have provided promising avenues to reliably integrate the molecular detail of biophysical process into the system behavior. Here, we review recent advances in the description of gene regulation as a system of biophysical processes that extend from specific protein-DNA interactions to the combinatorial assembly of nucleoprotein complexes. There is now basic mechanistic understanding on how promoters controlled by multiple, local and distal, DNA binding sites for transcription factors can actively control transcriptional noise, cell-to-cell variability, and other properties of gene regulation, including precision and flexibility of the transcriptional responses.**

[*] Correspondence: j.vilar@ikerbasque.org or lsaiz@ucdavis.edu

# Introduction

The process that leads to functional RNA and protein molecules from the information encoded in genes is known as gene expression. It starts with the binding of the RNA polymerase (RNAP) to the promoter, continues with transcription of the gene into RNA, and often concludes with translation into protein (1). This simple description is just the backbone of a much more complex set of events involving many processes that actively regulate, complement, affect, and critically refine all these three steps (2).

The complexity of gene expression is already evident at the very early stages of the process. The RNAP rarely just binds to DNA and starts transcription but there are molecules, such as transcription factors (TFs), that enhance, stabilize, hinder, and prevent the binding of the RNAP to the promoter (3). This local layer of control is embedded in the underlying dynamic organization of the genome, which determines to a large extent the accessibility of the RNAP to the promoter and to the information content of sets of genes that are spatially in the same region (4-8). In addition, the RNAP is not a simple molecule but a multi-subunit complex that, especially in eukaryotes, does not necessarily need to come pre-assembled to the promoter region or to be ready to start transcription upon binding (9, 10). Along the way, there are molecular mechanisms that affect RNA stability and its information content, such as alternative splicing and RNA editing (1). To close up the loop, proteins and functional RNA are in charge of orchestrating all these processes, thus regulating their own synthesis.

This short review focuses on key, well-characterized guiding principles that allow the description of gene expression in terms of systems of biophysical processes and on the application of these principles to actual systems, exemplified by the *lac* operon in prokaryotes and the retinoid X receptor in eukaryotes, which are both amenable to concise informative mechanistic descriptions. The goal is to accurately capture the effects of molecular interactions across scales up to the system behavior. To do so effectively, we will emphasize approaches that are scalable; namely, approaches that can be used with small and large systems, that can incorporate complex phenomena such as DNA looping, that use as few free parameters as possible, and that their molecular parameters can be inferred from the experimental data and reused in modeling subsequent experiments.

There are two types of important situations that we will not consider explicitly because of space limitations. One type includes elementary mechanisms, such a cooperative interactions, that are described in virtually any biochemistry and molecular biology textbook (1) and are applied to gene regulation exactly as described in the textbooks (11-14). The other type includes complex situations with missing key mechanistic information, such as eukaryotic enhancers (15, 16), which would extend the discussion to cover many potential mechanisms that are compatible with the observed experimental data. The most effective avenue to model this type of complex problems so far has been to supplement known biophysical mechanisms with phenomenological rules and assumptions (17, 18).

There are also many important aspects of gene expression that we will not be able to reach, including the effects of focused and dispersed transcription initiation (19), transcription elongation regulation (20, 21), transcriptional traffic (22), and translation regulation by microRNAs (23), to mention just a few. The general principles reviewed here also apply to a large extent to those situations.

## Modeling gene expression

A common starting point for most quantitative approaches to gene expression is a description based on reactions among molecular species (24-30). This description considers that there is a set of $i$ different transcriptional states $\delta_i$ and that for each of these states there is a given transcription rate $\Gamma_i$ that leads to mRNA, $m$. The simplest case with a single state would be a constitutive promoter with a constant transcription rate. The next step in complexity, a promoter with two states, already includes the potential for regulation, as for instance when a repressor turns off transcription upon binding the promoter. In general, the transitions between transcriptional states $\delta_i$ and $\delta_j$ with rates $k_{ij}$ depend on the numbers of the different molecular species of the system. For each mRNA molecule, proteins $p$ are produced at a rate $\Omega$. Typically, mRNA and proteins are degraded at rates $\gamma_m$ and $\gamma_p$, respectively. These reactions can be summarized as

$$
\begin{aligned}
\delta_i &\xrightarrow{k_{ij}} \delta_j \\
\delta_i &\xrightarrow{\Gamma_i} \delta_i + m \\
m &\xrightarrow{\gamma_m} \varnothing \\
m &\xrightarrow{\Omega} m + p \\
p &\xrightarrow{\gamma_p} \varnothing
\end{aligned}
\qquad (1)
$$

The advantage of using such an approach is that it allows a direct connection of the description parameters with biophysical properties such as free energies of binding and DNA elastic properties.

## The deterministic approach

When fluctuations are not relevant, either because they are small or because they can be averaged out (31), the set of Eqs. 1 is usually written in terms of concentrations using traditional deterministic rate equations:

$$
\begin{aligned}
\frac{dP_i}{dt} &= \sum_j (k_{ji} P_j - k_{ij} P_i), \\
\frac{d[m]}{dt} &= \sum_i (\Gamma_i / V_c) P_i - \gamma_m [m], \\
\frac{d[p]}{dt} &= \Omega [m] - \gamma_p [p].
\end{aligned}
\qquad (2)
$$

Here, $V_c$ is the reaction volume, $P_i = <\delta_i>$ is the probability of having the system in the transcriptional state $i$, $[m] = <m>/V_c$ is the mRNA concentration, and $[p] = <p>/V_c$ is the average protein concentration, with angular brackets $<...>$ representing averages.

The previous set of equations can be solved to obtain the steady state protein content concentration $[p]^{ss}$ as

$$[p]^{ss} = p_{max} \sum_i \chi_i P_i \ , \qquad (3)$$

where $p_{max} = \Gamma_{max}\Omega/\gamma_m\gamma_p V_c$ is the maximum concentration, $\Gamma_{max}$ is the maximum transcriptional activity, and $\chi_i = \Gamma_i/\Gamma_{max}$ is the normalized transcriptional activity. This result is extremely important because, besides collapsing the effects of many processes into a single parameter $p_{max}$, it directly connects microscopic probabilities of the transcriptional states with experimentally measurable quantities.

The deterministic approach, also known as mean-field approach, has been very useful to study systems with large numbers of molecules and negligible fluctuations. In the presence of fluctuations of small numbers of molecules, the average behavior of the system is still correctly described by the set of Eqs. 2. The applicability of the deterministic approach, however, could break down with the additional presence of nonlinear terms. The reason is that the average of nonlinear terms cannot generally be expressed in terms of concentrations. For instance, the kinetics of dimerization of a protein $p$ would involve the term $<p^2>$, which is not equivalent to $<p>^2 = (V_c[p])^2$. In general, the validity of the deterministic approach should be carefully assessed on a case by case basis, taking into account that neither small numbers of molecules nor nonlinear terms by themselves always prevent its applicability, as illustrated by genetic nonlinear oscillators that can function in the deterministic regime even with just a few mRNA molecules per cell (32).

## Control of gene expression

Control of gene expression is achieved through the dependence of the probability of the transcriptional states on the specific pattern of TFs that are assembled on DNA, as for instance, binding of an activator and absence of a repressor (2). In most instances, these interactions take place under quasi-equilibrium conditions and statistical thermodynamics can be used to express the probabilities of the states in terms of standard free energies and concentrations of the different regulatory molecules involved (33-35). The validity of the quasi-equilibrium assumption requires the binding kinetics, which by itself would be a completely reversible reaction, to be much faster than other cellular processes, such as cell growth, that could affect the binding process.

The key quantity in the thermodynamic approach is the statistical weight, or Boltzmann factor, which is defined in terms of the free energy $\Delta G_i$ of the state $i$ as $Z_i = e^{-\Delta G_i/RT}$. Its main feature is its proportionality to the probability of the state $i$,

$$P_i = e^{-\Delta G_i/RT}/Z \ . \qquad (4)$$

The normalization factor $Z = \sum_i Z_i$ is known as the partition function and the term $RT$ is, as usual, the gas constant, $R$, times the absolute temperature, $T$. This expression is particularly important because it encapsulates the dependence of the probabilities on the different molecular concentrations of regulatory molecules $[p_j]$ through

$$\Delta G_i = \Delta G_i^o - \sum_j d(i,j) RT \ln[p_j], \qquad (5)$$

where the terms $d(i,j)$ correspond to the number of molecules of the species $j$ in the state $i$ and $\Delta G_i^o$ is the corresponding standard free energy at 1M concentration. Therefore, if the free energies, or alternatively the probabilities, of the different states are known for given values of the concentrations of regulatory molecules, it is possible to obtain the probabilities for any concentration using the previous two equations. These two equations can be combined into

$$P_i = (\prod_j [p_j]^{d(i,j)}) e^{-\Delta G_i^o / RT} / Z, \qquad (6)$$

which has been a cornerstone in modeling quantitatively gene expression since the beginning of the field (30, 34). Its main advantage is that it only requires the values $\Delta G_i^o$ for each transcriptional state, which has traditionally been written on a table along with a description of the molecular configuration (34).

## Combinatorial complexity

The main advantage of using a free energy value for each transcriptional state may turn increasingly fast into a disadvantage when the number of components of the system increases. The reason is that there are potentially as many states as the number of possible ways of arranging the regulatory molecules on DNA, which grows exponentially with the number of components. The resulting combinatorial explosion in the number of states makes the straightforward application of Eq. 6 impracticable for systems with more than just a handful of components.

Several general approaches have been developed to tackle this exponentially large multiplicity in the number of states. They involve a diversity of methodologies that range from stochastic configuration sampling (36) to automatic generation of all the underlying equations (37). The complexity of the general problem makes each of these approaches work efficiently only on a particular type of problems, be it conformational changes, multi-site phosphorylation, or oligomerization (38-42).

In the case of gene regulation, it has been possible to capitalize on the unambiguous modular structure that macromolecular complexes typically have on DNA to capture this complexity in simple terms (43). The key idea is to describe the specific configuration, or state of the protein-DNA complex, through a set of $M$ state variables, denoted by $s = (s_1, ... s_k, ... s_M)$, which indicate whether a particular molecular component or conformation is present ($s_k = 1$) or absent ($s_k = 0$) at a specific position within the complex (43). The main advantage is that the free energy $\Delta G_i = \Delta G(s)$ and transcription rates $\Gamma_i = \Gamma(s)$ for each state can be specified as function of the state variables without explicitly enumerating all the states.

## The *lac* operon

The *E. coli lac* operon is the genetic system that regulates and produces the enzymes needed to metabolize lactose (44, 45). Besides opening the doors to the field of gene regulation, the *lac* operon has provided an example of a sophisticated regulation mechanism where all the components are known in great detail (46-52).

The main player in the control of transcription is the tetrameric *lac* repressor. In the absence of allolactose, a derivative of lactose, the *lac* repressor can bind to the main operator to prevent the RNAP from binding to the promoter and transcribing the genes. Binding of allolactose to the repressor substantially reduces its specific binding for the operator and transcription is de-repressed. The effects of the *lac* repressor on transcription are characterized as negative control. There is also positive control through the catabolite activator protein (CAP), which acts as an activator of transcription when glucose is not present by stabilizing the binding of the RNAP to the promoter.

This account of positive and negative control does not offer the whole picture of the underlying complexity. There are also two additional auxiliary operators that bind the repressor without preventing transcription (Figure 1A). Early on, they were considered just remnants of evolution because they bind the repressor very weakly and because elimination of either one of them has only minor effects in transcription. It was later observed that the simultaneous elimination of both auxiliary operators reduces the repression level by about 100 times (46-48). The reason for this astonishing effect is that the *lac* repressor can bind simultaneously two operators and loop the intervening DNA (Figure 1B). Thus, the main operator and at least one auxiliary operator are needed to form DNA loops that substantially increase the repressor's ability to bind the main operator. Without quantitative approaches, however, it is difficult to fully grasp how such weak auxiliary sites, as much as 300 times weaker in terms of binding affinity than the main operator, can help the binding so much.

To illustrate the important effects of the presence of the auxiliary operators, we examine in detail the case with two operators, the main operator, $O_m$, and an auxiliary operator, $O_a$. The main operator is located at the position of $O_1$ and the auxiliary operator is located at the position of either $O_2$ or $O_3$ (Figure 1A).

The key piece of information that allowed capturing the effects of DNA looping in quantitative detail was shown to be the decomposition of the free energy of the looped protein–DNA complex, $\Delta G^o_{l-c}$, into different modular contributions that take into account the binding to each operator and the looping contribution (49). Explicitly, $\Delta G^o_{l-c} = g_m + g_a + g_L$, where $g_m$ and $g_a$ are the standard free energy of binding to $O_m$ and $O_a$, respectively, and $g_L$ is the free energy of looping (Figure 1B).

This segmentation of the free energy allows for an efficient representation of all the transcriptional states in terms of state variables. These variables comprise $s_m$ and $s_a$, which indicate whether ($=1$) or not ($=0$) a repressor is bound to the main and auxiliary operator, respectively, and $s_L$, which indicates whether ($=1$) or not ($=0$) DNA forms the loop $O_m$-$O_a$.

The free energy of the system in terms of these three state variables is given by
$$\Delta G(s) = (g_m - RT \ln[n])s_m + (g_a - RT \ln[n])s_a \\ + (g_L + RT \ln[n])s_m s_a s_L + \infty(1 - s_m s_a)s_L, \quad (7)$$

where $[n]$ is the concentration of the *lac* repressor. The first two terms in the expression take into account the repressor binding to $O_m$ or $O_a$. The fourth term indicates that the presence of looping needs both operators occupied by a repressor; otherwise, the free energy would be infinite. Finally, the first three terms all together represent the binding of the repressor when the three state variables are equal to 1, which indicates that a single repressor is bound simultaneously to $O_a$ and $O_m$ and that there is a looping contribution to the free energy.

The normalized transcriptional activity is expressed in terms of state variables as
$$\chi(s) = (1-s_m)(\chi_a s_a + 1 - s_a). \tag{8}$$
This expression specifies that there is no transcription when the repressor is bound to $O_m$. When $O_m$ is free, transcription occurs at a maximum rate if $O_a$ is free and at rate $\chi_a$ if the repressor occupies $O_a$.

The advantage of using state variables is that Eqs. 7 and 8 completely specify the transcriptional properties of the *lac* operon. The steady state protein production is computed directly from $[p]^{ss} = p_{\max} \sum_s \chi(s) P(s)$, where the sum can be performed by hand or automatically using software like CplexA (53, 54). The resulting repression level is given succinctly by
$$\frac{p_{\max}}{[p]^{ss}} = \frac{(e^{g_m/RT} + [n])(e^{g_a/RT} + [n]) + [n]e^{-g_L/RT}}{e^{g_m/RT}(e^{g_a/RT} + [n]\chi_a)}. \tag{9}$$

This expression is important because it connects macroscopically measurable quantities, such as protein content in a cell population, with microscopic binding parameters. The value of $p_{\max}$ can be obtained from measurements for strains without repressor, which transcribe the *lac* genes at a maximum rate, and it is customary to report just the ratio $p_{\max}/[p]^{ss}$, which is known as repression level. In the case of the *lac* operon, all the parameters needed for modeling can be inferred from the experimentally available data.

For instance, when the auxiliary operator is deleted, or more precisely when it is mutated so that the binding is very low ($g_a \to \infty$), the repression level reduces to
$$p_{\max}/[p]^{ss} = 1 + [n]e^{-g_m/RT}. \tag{10}$$
From this expression and the data of experimental setups that used the sequence of $O_1$, $O_2$, and $O_3$ as a main operator, it is possible to obtain the free energy of binding for each operator (Figure 1C), which can be reused in subsequent modeling.

In the case of the $O_1$-$O_2$ loop for different sequences of the main operator, the only additional parameter needed to accurately reproduce the experimental data is the free energy of looping $g_L$ (Figure 1D). In this case, binding of the repressor to the auxiliary operator $O_2$ does not affect transcription and $\chi_a = 1$.

In the case of the $O_3$-$O_1$ loop, binding of the repressor to the auxiliary operator prevents CAP from activating transcription and the transcription rate is reduced to $\chi_a = 0.03$. In this case as well, just a single additional parameter for the free energy of looping $g_L$ is needed to reproduce most of the experimental data (Figure 1E). It turns out, however, that the deletion of $O_1$ in the strain labeled $O_3$-$O_{1X}$-X is not complete and the site is still able to form the $O_3$-$O_{1X}$ loop even though its binding is reduced by 5.5 kcal/mol, a factor 10,000 in terms of binding affinity. This moderate decrease in binding can be inferred from the Position Weight Matrix score of the specific sequence of the incomplete deletion (55). The computed repression level for the complete deletion of $O_1$, strain $O_3$-X-X, is shown as discontinuous line clearly below the incomplete deletion.

## The retinoid X receptor

Gene expression in eukaryotes is substantially more involved than in prokaryotes (2, 3, 56). Just the core of the eukaryotic transcriptional machinery itself involves a wide variety of components with oscillatory patterns of macromolecular assembly and phosphorylation (9, 57). In addition, there are many additional layers of control that extend from the accessibility and assembly of the transcriptional machinery at the promoter to the intracellular transport and regulation of mRNA and proteins. Despite all these differences, it has been argued that there are many general principles that apply to both prokaryotes and eukaryotes (2). We use the retinoid X receptor (RXR) to illustrate how the main ideas and methodology used in the *lac* operon can also be applied to this complex eukaryotic system.

RXR is a nuclear receptor that is responsible for regulating a large number of genes. It exerts its function by binding to DNA as homodimer, homotetramer, or obligatory heterodimerization partner for other nuclear receptors (58).

Similarly to the *lac* operon, RXR can bind multiple sites simultaneously as a tetramer by looping the intervening DNA (Figure 2A). A distinct feature, however, is that in the case of RXR, tetramers and dimers coexist in the cell and their relative populations are regulated by the RXR cognate ligands, which prevent the formation of tetramers besides imparting RXR the ability to recruit co-activators of transcription.

The first step in the signaling cascade for sensing the ligand concentration is regulation of the relative abundance of the oligomerization states of the RXR, which include tetramers, $n_4$, dimers, $n_2$, and non-tetramerizing dimers, $n_2^*$. The effects of the ligand are quantitated in general through the modulator function $f([l]) = [n_2^*]/[n_2]$, which describes the partitioning into the tetramerizing and non-tetramerizing dimers by the ligand $l$. In this system, the canonical ligand is the hormone 9cRA (9-*cis*-retinoic acid), a derivative of Vitamin A, which binds each RXR monomeric subunit independently of its oligomerization state (59) and prevents dimers with their two subunits occupied from tetramerazing (60). Therefore, considering $[n_2^*]$ as the concentration of dimers with two ligands bound and $[n_2]$ as the concentration of dimers with one or zero ligand leads to $f([l]) = [l]^2/(K_{lig}^2 + 2K_{lig}[l])$, where $K_{lig}$ is the ligand-RXR dissociation constant and $[l]$ is the concentration of the ligand (61). This process determines dimer and tetramer concentrations, which are related to each other through $[n_2]^2/[n_4] = K_{td}$, where $K_{td}$ is the tetramer-dimer dissociation constant.

Control of gene expression results from the dependence of the transcriptional response on the type of oligomeric species that are assembled on DNA (62). There are two differentiated types of responses (Figure 2B). The first type, referred to as response R1, involves a tetramer that simultaneously binds two non-adjacent DNA sites. Upon binding, the tetramer can bring a distal enhancer close to promoter region by looping DNA and control transcription. In this case, dimers do not elicit transcriptional responses. In general, promoting and preventing DNA looping has been found to be a fundamental mechanism for controlling the effects of distal enhancers (63, 64). The second type, denoted response R2, relies on differentiated recruitment abilities by different oligomerization states. Specifically, dimers can recruit a coactivator by binding of a region that is secluded in the tetramer (61).

The different configurations for binding of RXR to two DNA sites are described by the state variables $s_{1t}$ and $s_{2t}$ that indicate whether ($=1$) or not ($=0$) a tetramer is bound to site 1 and 2, respectively; $s_L$ that indicates whether ($=1$) or not ($=0$) DNA forms the loop between these two sites; and two additional state variables $s_{1d}$ and $s_{2d}$ that indicate whether ($=1$) or not ($=0$) a dimer is bound to site 1 and 2, respectively.

The free energy of the system in terms of these state variables is given by

$$\Delta G(s) = (g_1 - RT\ln[n_4])s_{1t} + (g_2 - RT\ln[n_4])s_{2t}$$
$$+ (g_L + RT\ln[n_4])s_{1t}s_{2t}s_L + \infty(1 - s_{1t}s_{2t})s_L$$
$$+ (g_2 - RT\ln([n_2]+[n_2^*]))s_{1d} + (g_2 - RT\ln([n_2]+[n_2^*]))s_{2d} \quad (11)$$
$$+ \infty(s_{1t}s_{1d} + s_{2t}s_{2d}).$$

Here, $g_1$ and $g_2$ are the standard free energies of binding to sites 1 and 2, respectively, which are assumed to be the same for all three oligomeric species, and $g_L$ is the free energy of looping. The first four terms of this expression are equivalent to those for the *lac* operon in Eq. 7 since it is the same type of tetrameric binding to two sites. The fifth and sixth terms represent the binding of a dimeric species to site 1 and 2, respectively. The last term indicates that dimers and tetramers cannot be bound simultaneously to the same site by assigning an infinite free energy to those states.

The normalized transcriptional activities for responses R1 and R2 are expressed in terms of state variables as

$$\chi_{R1}(s) = \chi_{ref} + (1-\chi_{ref})s_L,$$
$$\chi_{R2}(s) = \chi_{ref} + \chi_d(s_{1d}+s_{2d}) + (\chi_{dd}-\chi_d)s_{1d}s_{2d}, \quad (12)$$

where $\chi_{ref}$ does not depend on the ligand concentration and is the normalized basal activity of the promoter in absence of any activation. The explicit forms of $\chi_d = (1-(1+[l]/K_{lig})^{-2})(1-\chi_{ref})$ and $\chi_{dd} = (1-(1+[l]/K_{lig})^{-4})(1-\chi_{ref})$ implement that at least one of the ligand-binding sites of one dimer and of a pair of dimers, respectively, needs to be occupied by the ligand for the coactivator to be recruited.

It is straightforward to obtain analytic expressions of the transcriptional activity from Eqs. 11 and 12 using software packages like CplexA (53) but it is more illustrative for the purposes of this review to focus on the functional regime that guarantees that there is response to changes in the ligand concentration.

The functional regime considers two properties. The first one is that the total RXR concentration is sufficiently high for it to significantly bind DNA. The second one is that the concentration of tetramers is low enough for them not to completely saturate the binding. The reason is that for typical values of $g_L$, tetramers bind more strongly to two DNA sites simultaneously than dimers do to a single DNA site, as in the case of the *lac* operon (43, 49, 65). Under these conditions the representative states, described by $s = (s_{1t}, s_{2t}, s_L, s_{1d}, s_{2d})$, are those with a tetramer bound to the two sites simultaneously, $s = (1,1,1,0,0)$, and with one dimer bound to each of the two sites, $s = (0,0,0,1,1)$. The corresponding statistical weights for these states, the only ones needed in this case, are $Z_{(1,1,1,0,0)} = [n_4]e^{-(g_1+g_2+g_L)/RT}$ and $Z_{(0,0,0,1,1)} = ([n_2]+[n_2^*])^2 e^{-(g_1+g_2)/RT}$, respectively.

The key implication of this regime is that the steady state protein production, computed from $[p]^{ss} = p_{max} \sum_s \chi(s) P(s)$, simplifies in such a way that the transcriptional responses are governed by the reduced expressions

$$[p]^{ss}_{R1} / p_{max,R1} = \chi_{ref} + (1-\chi_{ref})P_t,$$
$$[p]^{ss}_{R2} / p_{max,R2} = \chi_{ref} + (1-(1+[l]/K_{lig})^{-4})(1-\chi_{ref})(1-P_t), \quad (13)$$

where

$$P_t = \frac{1}{1+\left(1+f([l])\right)^2 e^{g_L/RT} K_{td}} \quad (14)$$

is the probability of the state $s = (1,1,1,0,0)$.

The particular form of $P_t$ is exceptionally remarkable because it imparts precision and flexibility to the transcriptional responses, two properties that are the cornerstone of natural gene expression systems but that have proved to be highly elusive because of their seemingly antagonistic character (66). Precision ensures that the transcriptional response is consistently triggered at a given ligand concentration irrespective of the particular total RXR concentration, which cancels out in the reduced equations that govern the system behavior. Flexibility, on the other hand, allows the precise triggering point to be altered both at the individual promoter level through $g_L$ (67, 68) and at a genome-wide scale through $f([l])$ and $K_{td}$.

To compare with the experimental data, the most convenient approach is to use the normalized fold induction (*NFI*), which is defined as $NFI = (FI-1)/(FI_{max}-1)$, where $FI$ is the fold induction and $FI_{max}$ is its maximum value. The value of $FI$ is obtained experimentally as the actual expression of a gene over its baseline expression and in mathematical terms as $FI = [p]^{ss}/(p_{max}\chi_{ref})$. In terms of the *NFI*, the results do not depend on parameters related to the baseline and maximum expression levels and it becomes possible to effectively compare experiments on different promoters and cell lines. Its explicit form for response R2 and R1 is

$$NFI_{R1} = P_t,$$
$$NFI_{R2} = \left(1-(1+[l]/K_{lig})^{-4}\right)(1-P_t), \quad (15)$$

respectively. Importantly, the only parameters needed to characterize the shape of the response in the functional regime are $K_{lig}$ and $K_{td}$, which have been measured experimentally, and $g_L$, which can be inferred by adjusting its value to reproduce the experimental data.

A fully predictive framework without free parameters has been obtained with this approach because it collapses most of the intracellular complexity into just one unknown parameter $g_L$. Therefore, once this parameter is known for a particular experimental set-up (specific cell type, cellular conditions, and promoter), it can be used to predict other responses just from thermodynamic principles.

One possibility is to use the value of $g_L$ inferred for one type of response to predict the other one. There is experimental data that tested in the same cell type and promoter both types of transcriptional responses, one mediated by an enhancer (response R1) and the other, by a coactivator (response R2). The results of the model indicate that just a single value of $g_L$ is needed to reproduce with high accuracy the experimental data in both cases (Figure 2C).

Another possibility is to use the value of $g_L$ inferred for one ligand to predict the response to other ligands. The all-*trans*-retinoic (atRA) was tested early on as a potential candidate for the RXR cognate ligand and was observed that binding was present, but it was very weak (69, 70). The values of $g_L$ inferred for 9cRA responses, can be used to closely match the experimental transcription data in response to atRA without any free parameter by just changing the value of the ligand-binding constant, $K_{lig}$, to the corresponding one for atRA (Figure 2D).

## Combinatorial assembly of nucleoprotein complexes

There are many situations in which the DNA loop is formed not by a single protein, as in the *lac* operon and RXR, but by a protein complex that is assembled on DNA as the loop forms. The term combinatorial assembly is used because there are many potential complexes that can arise from the combinations of binding to multiple sites, even when just a single TF is involved. An illustrative example is present in the regulation of phage $\lambda$. It has two operators located 2.4 kb away from one another and each operator contains a tandem of three sites where phage $\lambda$ cI repressors can bind as dimers. In this case, two dimers bound to an operator can form an octamer with two dimers bound to another operator by looping the intervening DNA (43, 71, 72). Another example is the interaction of TFs bound at distal enhancers with the transcriptional complexes bound at the promoter (63). To study this type of problems, it is crucial to properly take into account that proteins bound to distal DNA regions can interact with each other only if DNA looping is present.

Interactions mediated by DNA looping would lead to terms with products of three or more state variables in the free energy (43). An illustrative example is $(g_L + \sum_{i,j} e_{i,j} s_{U,i} s_{D,j}) s_L$, where the state variables $s_{U,i}$, $s_{D,j}$, and $s_L$ indicate whether ($=1$) or not ($=0$) a protein is bound to site $i$ at the upstream DNA region, a protein is bound to site $j$ at the downstream DNA region, and DNA looping is present, respectively. The quantities $e_{i,j}$ account for the interactions between proteins bound at different DNA regions and $g_L$ is the free energy of looping. The formation of the DNA loop would be energetically favorable only when a sufficient number of interactions can be achieved between the two DNA regions. In the case of phage $\lambda$, only octamers and dodecamers are able to form the looped complex among the many possible combinations of binding (43, 71, 72). In turn, the presence of DNA looping can enhance DNA binding through the interactions that can be established between the two DNA regions, which can lead to highly cooperative phenomena in the formation of the nucleoprotein complex (42).

## Stochastic kinetics

The *lac* operon and RXR have been used so far in this review to demonstrate how biophysical principles can be used to efficiently capture the system behavior when noise in the form of random fluctuations is not relevant. The very same principles can be extended to take into account the inherent stochastic nature of the underlying processes in a wide range of situations. An efficient avenue to do so is to consider the dynamics of the macromolecular complexes that control gene expression through the stochastic dynamics of the state variables (43, 73).

The dynamics of the macromolecular complex can be described in terms of components that can change in a transition. For the widespread case in which only one component can change at a given time, either the component $i$ gets into or out of the complex, one can define *on* ($k_{on}^i$) and *off* ($k_{off}^i$) rates for the association-like and dissociation-like rates, respectively, which in general depend on the pre-transition and post-transition states of the complex.

The explicit dynamics can be obtained by considering the change in state variables as reactions given by

$$s_i \xrightarrow{r_i} (1-s_i) \text{, with } r_i = (1-s_i)k_{on}^i(s) + s_i k_{off}^i(s). \tag{16}$$

These reactions change the variable $s_i$ to 1 when it is 0 and to 0 when it is 1, representing that the element gets into or out of the complex. Typically, the *on* rate does not depend as strongly on the state of the complex as the *off* rate. The *on* rate is essentially the rate of transferring the component from solution to the complex. The *off* rate, in contrast, depends exponentially on the free energy change.

The principle of detailed balance (33) can be used to obtain the *off* rates from the *on* rates:

$$k_{off}^i(s) = k_{on}^i(s')e^{-(\Delta G(s')-\Delta G(s))/RT}. \tag{17}$$

The remarkable property of this expression is that reactions with known rates can be used to infer the rates of more complex reactions from the equilibrium properties; for instance, to infer dissociation rates for different binding sites from a single association rate (43). In general, the association rate could also depend on the state of the complex and its free energy, as for instance if the presence of a TF facilitates the association of another TF. If this dependence is included in the *on* rate, Eq. 17 can also be applied straightforwardly to obtain the *off* rate.

The stochastic dynamics of the resulting networks of reactions and transitions can then be obtained with kinetic Monte-Carlo simulations using well-established algorithms (26, 74, 75).

## Noise and fluctuations in the *lac* operon

Stochastic effects in the *lac* operon have been known to be important since the late 1950s (76), predating the discovery of gene regulation (45). The most salient example is the all-or-none induction process (76), which was measured at the single-cell level with a resolution of a few molecules of the gene products per cell (77). This effect has its roots in the amplification of the inherent stochastic fluctuations of transcription and translation processes (78-81) close to the boundary that separates the induced from non-induced states of the *lac* operon (24, 82).

The underlying molecular mechanisms and parameters have been shown to shape transcriptional noise to a large extent (43, 49, 83-86). To illustrate these effects in the *lac* operon, we discuss first regulation through just the main operator. The use of state variables leads to a single reaction that describes both the binding and unbinding of the repressor to the main operator:

$$s_m \xrightarrow{r_m} (1 - s_m), \text{ with } r_m = [n]k_a\left((1-s_m) + s_m e^{g_m/RT}/[n]\right). \quad (18)$$

Here, the *on* rate is given by $[n]k_a$, where $k_a$ is the association rate constant, and the *off* rate, $k_a e^{g_m/RT}$, is obtained from the detailed balance principle. The transcription rate is described by

$$m \xrightarrow{\Gamma_S} m+1, \text{ with } \Gamma_S = \Gamma_{max}(1-s_m) \quad (19)$$

and mRNA degradation, protein production, and protein degradation are described by the stochastic counterpart of Eqs. 1. The time courses of the number of proteins produced from this promoter show relatively small fluctuations for the experimental values of the parameters (Figure 3A). The downside of having just a binding site for regulation is that repression is relatively weak and a substantial number of proteins are produced.

To increase repression, there exist two simple alternatives. The first one is to consider a stronger site. For a site 50 times stronger than the wild-type main operator, protein production would be close to the value expected for the *lac* operon with the three operators. In this case, the average protein production is reduced about 50 times, as expected from the deterministic theory, but fluctuations increase dramatically (Figure 3B). There are infrequent mRNA bursts that lead to large protein amounts that decay in a few hours and long periods of time without any protein at all. The second alternative is to include more repressors. For a repressor concentration 50 times higher than in wild-type, the average protein production is reduced about 50 times and the fluctuations remain relatively small (Figure 3C). In this case, mRNA production happens in smaller quantities but more frequently. The physiological downside is that the repressor production would have to be 50 times higher than in wild-type and if that happens for all the proteins of the cell, *E. coli* would have to be 50 times more crowded.

A more efficient alternative to increase repression is to use DNA looping, which has been chosen by evolution not only in the *lac* operon but also in a large variety of systems. The computational approach in this case is slightly more involved because it has to take into account that an operator can be bound by a repressor in solution or by a repressor bound to the other operator thus forming a DNA loop (Figure 1B). For binding to the main operator, these two processes are represented by

$$s_m \xrightarrow{r_m} (1 - s_m), \text{ with } r_m = [n]k_a\left((1-s_m) + s_m(1-s_a s_L) e^{g_m/RT}/[n]\right), \quad (20)$$

$$\{s_L, s_m\} \xrightarrow{r_{L,m}} \{1-s_L, 1-s_m\}, \text{ with } r_{L,m} = e^{-g_L/RT} k_a s_a \left((1-s_L)(1-s_m) + s_L s_m e^{(g_m+g_L)/RT}\right). (21)$$

For the binding to the auxiliary operator, the reactions have the same representation except that the terms $s_m$, $s_a$, and $g_m$ are replaced by $s_a$, $s_m$, and $g_a$, respectively.

The stochastic kinetics of the regulation through the $O_1$-$O_2$ loop shows a small average number of proteins with low fluctuations, thus behaving in a very similar way as a single operator with 50 times more repressor (Figure 3D). Therefore, DNA looping in this case allows the system to achieve the same behavior as it would with 50 times more repressors.

Intuitively, both looping and high repressor concentration lead to lower noise than a single strong site because of the characteristic time scales involved. In the strong site case, there are long periods of time with maximum transcriptional activity and long periods without any activity, which results in the number of proteins fluctuating strongly between high and low values. In the cases of looping and high repressor concentration, the *off* rate of the repressor from the main operator is 50 times larger than for the strong site and the average *on* rate increases accordingly to keep the same repression level. Therefore, the switching between transcriptional states is very fast and mRNA production is in the form of short and frequent bursts. This lack of long periods of time with either full or null production gives a narrower distribution of the number of proteins. Explicitly, the coefficient of variation of protein (mRNA) content shown in Figure 3 for the strong site, high repressor concentration, and DNA looping cases is 2.3 (12.9), 0.81 (4.8), and 0.95 (5.4), respectively.

## Discussion

Gene expression relies on intricate molecular mechanisms to function in extraordinarily diverse intra- and extra-cellular environments. Biophysical approaches have provided new avenues to unravel how these different levels of molecular complexity contribute to the observed behavior. The results reviewed here show that the underlying complexity of biological systems is not just an accident of evolution but has a functional role.

Explicitly, the *lac* operon exemplifies how escalating complexity from one to two operators introduces stronger repression while preserving low transcriptional noise, which is not possible with a stronger single binding site.

In the case of the RXR, the additional complexity embedded in the control of its oligomeric state by the cognate ligand and its ability to bind simultaneously single and multiple DNA sites has been shown to impart precision and flexibility, two seemingly antagonistic properties, to the sensing of cellular signals.

This type of regulated oligomerization has also been observed explicitly in other transcription factors that can bind multiple DNA sites simultaneously, such as the tumor suppressor p53 (87), the nuclear factor κB (NF-κB) (88, 89), the signal transducers and activators of transcription (STATs) (90), and the octamer-binding proteins (Oct) (91, 92). In these systems, the properties of self-assembly, and the partitioning into low and high order oligomeric species, are strongly regulated and modulated by several types of signals, such as ligand binding (60), protein binding (93, 94), acetylation (95), and phosphorylation (92, 96).

The combined presence of flexibility and precision in the control of gene expression, as explicitly shown for RXR, allows a single TF to simultaneously regulate multiple genes with promoter-tailored dose-response curves that consistently maintain their diverse shapes for a broad range of the TF concentration changes.

Thus, the complexity of multiple repeated distal DNA binding sites both in prokaryotes and eukaryotes, far from being just a remnant of evolution or a backup system as often assumed, can confer fundamental properties that are not present in simpler setups.

## Acknowledgments

This work was supported by the MINECO under grants FIS2009-10352 and FIS2012-38105 (J.M.G.V.) and the University of California, Davis (L.S.).

# Figure legends

**Figure 1:** Gene expression in the *lac* operon for different operator configurations. **(A)** The relative positions of the main operator $O_1$ and the auxiliary operators $O_2$ and $O_3$ are shown as black rectangles on the black line representing DNA. The binding site for CAP is shown as a gray rectangle. **(B)** A representation of the *lac* repressor is shown looping DNA (black line) bound to the main and auxiliary operators (white rectangles). The contributions to the standard free energy of the looped DNA-repressor state from binding the main and auxiliary operators and from looping DNA are indicated by $g_m$, $g_a$, and $g_L$, respectively. **(C-E)** The repression level computed from Eq. 9 (lines) and experimentally measured by Oehler et al. (46) (symbols) are shown for different combinations of operator replacements and deletions, including configurations with just the main operator, the $O_1$-$O_2$ loop, and the $O_3$-$O_1$ loop. The notation on each curve indicates the operator sequence at the positions of the operators $O_3$-$O_1$-$O_2$. A complete deletion is indicated by X and the partial deletion of $O_1$ is indicated by $O_{1X}$. The values of $g_1$, $g_2$, $g_3$, and $g_{1X}$ indicate the standard free energy of binding (in kcal/mol) to the sequences of $O_1$, $O_2$, $O_3$, and $O_{1X}$, respectively. The value of $g_L$ used is shown (in kcal/mol) for each type of loop.

**Figure 2:** RXR-mediated transcriptional responses to 9cRA and atRA ligands. **(A)** A prototypical arrangement of binding sites for RXR and an enhancer element are shown as black and grey rectangles, respectively, on a black line representing DNA. **(B)** In response R1, an RXR tetramer loops DNA (represented as a continuous line) to bring an enhancer close to the promoter region. In response R2, an RXR dimer recruits a coactivator to the promoter region. **(C, D)** The normalized fold induction for responses R1 and R2 (lines) was computed from Eqs. 15 with the experimental values $K_{lig} = 8$ nM for 9cRA (97) or $K_{lig} = 350$ nM for atRA (98), and $K_{td} = 4.4$ nM (99). The value of the free energy of looping $g_L$ (shown in kcal/mol) depends on the specific promoter and cell line. **(C)** Response to 9cRA for a promoter incorporating two RXR binding sites with (left) and without (right) a distal enhancer, which considers responses R1 and R2, respectively. Experimental gene expression data (symbols) was obtained from Yasmin et al. (62). **(D)** Responses to 9cRA and atRA for promoters without a distal enhancer (response R2). Experimental gene expression data (symbols) correspond to reporter plasmids ADH-CRBPII-LUC (left) and TK-CRBPII-LUC (right) from Heyman *et al.* (70).

**Figure 3:** Transcriptional noise in the *lac* operon. **(A, B, C)** Time courses of the number of protein and mRNA (shown as negative values) produced from a promoter with just the main operator described by Eqs. 18 and 19 and the stochastic implementation of Eqs. 1. The values of the common parameters for all three panels are $\Gamma_{max} = 0.5$ s$^{-1}$, $\Omega = 0.01$ s$^{-1}$, $\gamma_m = 3.3 \times 10^{-3}$ s$^{-1}$, $\gamma_p = 9.2 \times 10^{-5}$ s$^{-1}$, and $k_a = 2.2 \times 10^{-6}$ M$^{-1}$s$^{-1}$. The values of the remaining parameters are **(A)** $g_m = -13.1$ kcal/mol and $[n] = 15$ nM for wild-type $O_m$ and wild-type repressor concentration; **(B)** $g_m = -13.1 - RT \ln 50$ kcal/mol and $[n] = 15$ nM for 50 times stronger Om and wild-type repressor concentration; **(C)** $g_m = -13.1$ kcal/mol and $[n] = 750$ nM for wild-type $O_m$ and 50 times more repressor. **(D)** Time courses of the number of protein and mRNA (shown as negative values) produced from a promoter with the operators $O_1$ and $O_2$ described by Eqs. 19, 20, and 21 and the stochastic implementation of Eqs. 1. The values of the parameters are the same as in panel **(A)** with the addition of $g_a = -11.6$ kcal/mol and $g_L = 8.30$ kcal/mol.

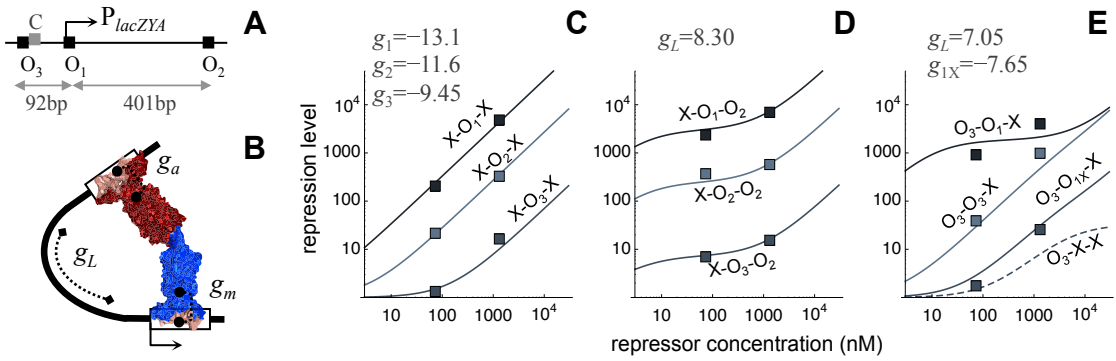

Figure 1

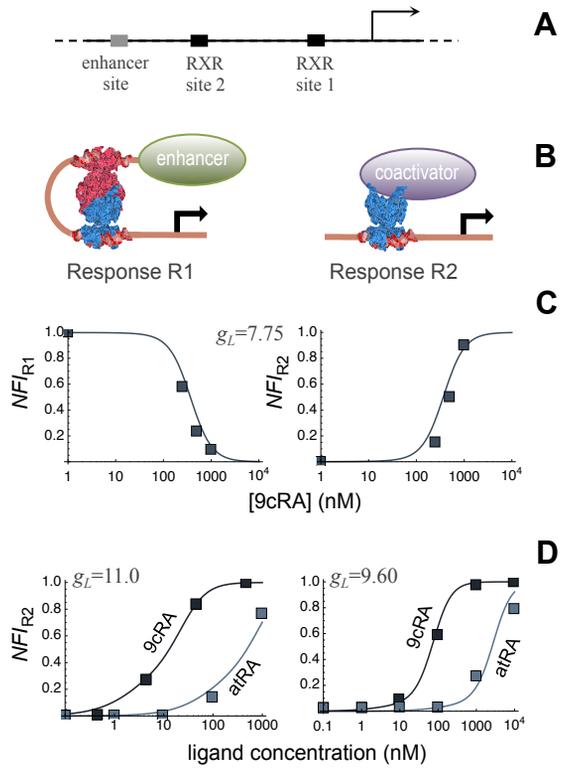

Figure 2

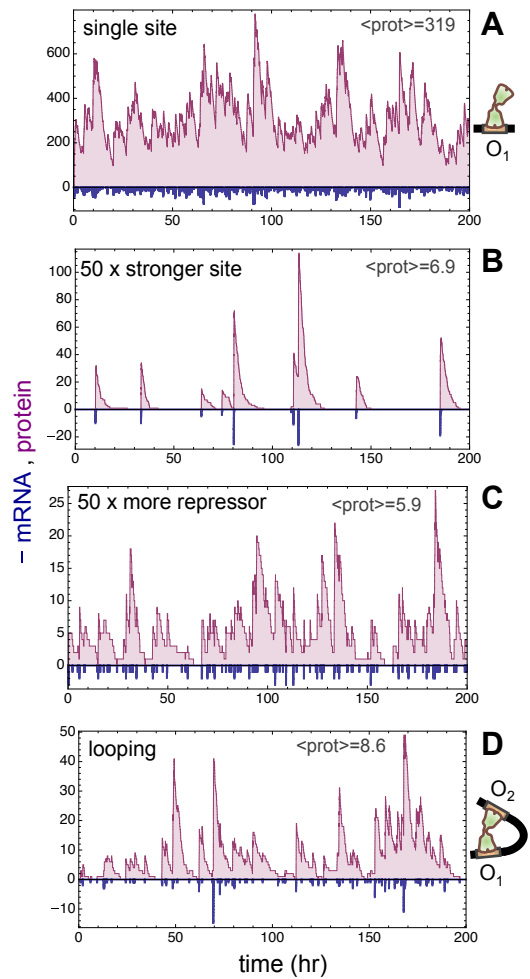

Figure 3